\documentclass[a4paper,12pt,final]{article}
\usepackage{amsmath,amsfonts,amssymb,amsthm,amstext,eucal}
\usepackage{graphicx,lscape}

\font\mybb=msbm10 at 10pt
\def\bb#1{\hbox{\mybb#1}}
\def\bR {\bb{R}}
\def\bZ {\bb{Z}}
\def\bE {\bb{E}}

\def\bC {\bb{C}}
\def\cM {{\cal M}}
\DeclareMathOperator{\dvol}{dvol}

\newcommand{\ISO}{\mathrm{ISO}}
\newcommand{\SO}{\mathrm{SO}}

\providecommand{\bysame}{\leavevmode\hbox to3em{\hrulefill}\thinspace}

\topmargin
-1.5cm
\textwidth
15.5cm
\textheight
24cm
\oddsidemargin
0.7cm
\evensidemargin
1.2cm

\begin{document}


\begin{titlepage}
\vfill
\begin{flushright}
CERN/2002-066\\
WIS/12/02-MAR-DPP\\
hep-th/0203243\\
\end{flushright}

\vfill
\begin{center}
\baselineskip=16pt
{\Large\bf M-theory lift of  brane-antibrane systems \\
and localised closed string tachyons}
\vskip 0.3cm
{\large {\sl }}
\vskip 10.mm
{\bf ~Ra\'ul Rabad\'an $^{*,1}$ ~and ~Joan Sim\'on 
$^{\dagger,2}$}\\
\vskip 1cm

{\small
$^*$
  Theory Division CERN,\\
  CH-1211 Gen\`eve 23, Switzerland}\\
\vskip.5cm
{\small
$^\dagger$ Department of Particle Physics, The Weizmann Institute 
of Science \\
Herzl Street 2, 76100 Rehovot, Israel}\\ 
\end{center}
\vfill
\par
\begin{abstract}
We discuss the lift of certain D6-antiD6-brane systems to M-theory. 
These are purely gravitational configurations with a bolt singularity. 
When reduced along a trivial circle, and for large bolt radius, the 
bolt is related to a non-supersymmetric orbifold type of singularity where 
some closed string tachyons are expected in the twisted sectors. This is a 
kind of open-closed string duality that relates open string tachyons on one 
side and localised tachyons in the other. We consider the evolution of the 
system of branes from the M-theory point of view. This evolution gives rise 
to a brane-antibrane annihilation on the brane side. On the gravity side, the 
evolution is related to a reduction of 
the order of the orbifold and to a contraction of the bolt to a nut or flat 
space if the system has non-vanishing or vanishing charge, respectively.  

We also consider the inverse process of reducing a non-supersymmetric 
orbifold to a D6-brane system. For $\bC^2/\bZ_N \times \bZ_M$, the reduced 
system is a fractional D6-brane at an orbifold singularity 
$\bC/\bZ_{\text{M}}$.

\end{abstract}
\begin{quote}

\vfill
 \hrule width 5.cm
\vskip 2.mm
{\small
\noindent $^1$ E-mail: Raul.Rabadan@cern.ch \\
\noindent $^2$ E-mail: jsimon@weizmann.ac.il\\
}
\end{quote}
\end{titlepage}

\setcounter{page}{1}

\section{Introduction}

It is well known that the lift to M-theory of a system of parallel D6-branes 
~\cite{hk85,t95} corresponds to a purely geometric background, the Taub-NUT 
metric. When the position of N of these D6-branes coincide, one gets an 
$A_{N-1}$ singularity at a point in the multi Taub-NUT space. In this paper,
we would like to make a step forward in the relation between the physics
of D6-branes at strong coupling and purely gravitational backgrounds in
eleven dimensional supergravity by studying the lift of a system of coincident
D6-$\overline{\text{D}6}$ branes to M-theory. We shall primarily be concerned 
with the geometry describing these configurations, its evolution as branes and 
antibranes annihilate each other\footnote{See, for instance, \cite{anni}.} and some similarities between the qualitative
patterns that we find in this evolution and some recent results on
the evolution due to the condensation of localised closed
string tachyons in non-supersymmetric orbifold singularities 
~\cite{orbi,aps01,v01,dv01,hkmm01,twisted}. 

In particular, we shall study the lift to M-theory of the generically non-BPS
configurations found in ~\cite{zz99,bmo00,lt01} preserving $\ISO(1,6)\times
\SO(3)$. The latter depend on three parameters. The subset of configurations 
in which we will be interested in corresponds to setting one of them 
to zero. These particular geometrical configurations look like 
$\bR^{1,6}\times\cM_4$, for some curved four dimensional manifold. It turns 
out that $\cM_4$ has a {\sl bolt} type singularity, that is, a locus
of conical singularities, whose conical defects depend on the mass and the
charge of the configuration. The brane-antibrane annihilation expected
in the open string description gives rise to a reduction in the size of
the bolt and a desingularization of the conical singularities, by which they
become ``less conical''. In the sector of non-vanishing charge, the bolt 
becomes a nut, whereas in the vanishing charge sector, the bolt disappears.

Locally, when the size of the bolt is big, the system looks like $\cM_4
\sim \bC\times\bC/\bZ_{\text{M}}$. The size of the bolt is proportional to 
the product of the number of branes, the number of antibranes and 
$(g_s l_s)^2$. Thus, big bolt limit means that $g_s^2 N\bar{N}$ is big, 
i.e. the number of branes and antibranes should be large in order to keep a 
small string coupling. Thus, by reducing along a trivial circle,
the original D6-$\overline{\text{D}6}$ system is related to a 
$\bC/\bZ_{\text{M}}$ orbifold in the forementioned limit. Whenever $M\neq 1$,
there are closed string tachyons in the twisted sectors. Recent studies
~\cite{aps01,v01,dv01} suggest that this system
evolves to flat space making the cone ``less conical'' by a sequence of
transitions
\[
  \bC/\bZ_{2l+1} \,\rightarrow\,\bC/\bZ_{2l-1} \,\dots \,\rightarrow\,
  \bC/\bZ_{2l^\prime-1} \quad (l^\prime< l)
\]
Our qualitative comparison in the large bolt limit suggests
a relation between brane-antibrane annihilation and twisted tachyon evolution. 
And in particular, each transition
$(l\to l-1)$, which reduces the order of the orbifold by two, is related to 
the annihilation of a D6-$\overline{\text{D}6}$ pair.

In the second part of the paper, and motivated by the previous relation,
we start from a non-supersymmetric orbifold acting on $\bC^2$ in type IIA,
lift the configuration to M-theory using a trivial transverse circle and 
reduce it along a non-trivial circle in $\bC^2$. One expects such system to be
the local description for an unstable system of branes. In particular, we
consider $\bC^2/\bZ_{\text{N}}\times \bZ_{\text{M}}$, where each abelian
group preserves different supersymmetry, so that the full orbifold is
non-supersymmetric. The interpretation of the reduced system is in terms
of fractional D6-branes living on a $\bC/\bZ_{\text{M}}$ singularity.
Here the closed string tachyonic instabilities cannot be mapped to open string tachyons as in the previous case.

The organisation of the paper is as follows. In section 2, we revisit the 
construction of supergravity solutions given in ~\cite{zz99,bmo00},
paying attention to the particular case of D6-$\overline{\text{D}6}$-branes.
These solutions depend on three parameters. We discuss the scaling limits
leading to BPS configurations, generalising the discussion in ~\cite{bmo00}.
We consider the lift of such configurations to M-theory
and argue why it is interesting for us to set one of the parameters to zero.
In this way, we get a two parameter family of solutions, where the parameters 
can be mapped to the Ramond-Ramond (RR) charge and mass of the system. 
In section 3, we analyse this solution in detail, both in the charged
and uncharged sectors. In section 4, we discuss the evolution of 
the system and we compare the open and closed string descriptions.
Section 5 is devoted to the study of the inverse problem: going from a 
non-supersymmetric orbifold to a local description of a system of D6-branes. 
In particular, we consider a $\bC^2/\bZ_{\text{N}} \times \bZ_{\text{M}}$ 
non-supersymmetric orbifold.

\section{From brane-antibranes to M-theory}

In ~\cite{zz99}, the most general solution to the supergravity equations
of motion with $\ISO(1,p)\times \SO(9-p)$ symmetry and carrying the
appropriate Ramond-Ramond (RR) charge was integrated. It was subsequently
interpreted in ~\cite{bmo00} as a system of coincident 
Dp-$\overline{\text{D}p}$ branes. In this work, we shall concentrate on 
the D6-$\overline{\text{D}6}$ system. 
In the Einstein frame, the configuration is described by

\begin{equation}
  \begin{aligned}
    g_E & =  e^{2A(r)} ds^2(\bE^{1,6}) + e^{2B(r)} (dr^2 + r^2 d\Omega^2_2) \\
    \Phi & =  \Phi(r) \\
    C_{(7)} & =   e^{\Lambda(r)} \dvol(\bE^{1,6})
  \end{aligned}
 \label{ansatzd6}
\end{equation}

where $g_E$ is the ten dimensional metric, $\Phi$ is the dilaton and 
$C_{(7)}$ is the RR seven form potential. The set of scalar functions
characterising the above configuration is given by

\begin{equation}
  \begin{aligned}
    A(r) & =  -\frac{3}{64}c_1\,h(r) 
    - \frac{1}{16} \log[\cosh(k h(r)) - c_2 \sinh(k h(r))] \\
    B(r) & = \log[f_-(r) f_+(r)] -7A(r) \\
    \Phi(r) & = c_1\,h(r) + 12\,A(r) \\
    e^{\Lambda(r)} & = - \sqrt{c_2^2 -1} 
    \frac{\sinh(k h(r))}{\cosh(k h(r)) - c_2 \sinh(k h(r))}
  \end{aligned}
 \label{setfunctions}
\end{equation}
where 

\begin{equation*}
  \begin{aligned}
    f_{\pm} & = 1 \pm \frac{r_0}{r} \\
    h(r) & = \log[f_-(r)/f_+(r)]  \\
    k &= \sqrt{4-\frac{7}{16}c_1^2} ~.
  \end{aligned}
\end{equation*}

Thus, it depends on two dimensionless parameters $\{c_1\,,c_2\}$ defined
in the ranges $c_2\geq 1$, $-\frac{8}{\sqrt{7}}\leq c_1\leq 0$, and a third
one $r_0$, with dimensions of length satisfying $r_0\geq 0$.

The charge (Q) and mass (M) of this solution were computed in
~\cite{bmo00} and we shall follow their conventions. They are expressed
in terms of $\{r_0\,,c_1\,,c_2\}$ as follows

\begin{eqnarray}
  Q & = & 2\,\text{P}\cdot k\,r_0\,\sqrt{c_2^2 -1} \label{charge} \\
  M & = & \text{P}\cdot r_0\left[2c_2\cdot k -\frac{3}{2}c_1\right] 
 \label{mass}
\end{eqnarray}

where $\text{P}=\frac{1}{16}\frac{V_6}{G^N_{10}}$, $V_6$ being the spacelike
volume spanned by the branes and $G^N_{10}$ stands for the ten dimensional
Newton's constant. Written in string units, 
$\text{P}=\frac{\pi}{(2\pi)^7}\frac{V_6}{g_s^2 l_s^8}$, where $g_s$ is the 
string coupling constant and $l_s$ is the string length 
$l_s^2 = \alpha^\prime$.

Notice that in general the configuration is non-BPS $(M\neq Q)$, as expected,
and it is useful to introduce the difference between these observables

\begin{equation}
  \delta M \equiv M - Q = \text{P}\,r_0
  \left[2k\left(c_2 - \sqrt{c_2^2 -1}\right) -
  \frac{3}{2}c_1\right] \,.
 \label{nonbps}
\end{equation}

\subsection{BPS limits}

The first natural question to address is how to recover the well-known
BPS configurations corresponding to N D6-branes 
(or $\overline{\text{D}}$6-branes)
from the general solution \eqref{ansatzd6}. At this point, we would like
to point out that there are more possibilities than the one discussed
in ~\cite{bmo00}. Indeed, the idea there was to take a certain scaling limit
in the set of parameters $\{r_0\,,c_1\,,c_2\}$, or equivalently in
$\{r_0\,,k\,,c_2\}$, such that the charge Q remains finite while 
$\delta M\to 0$. As discussed in ~\cite{bmo00}, one possibility is to consider

\begin{equation}
  r_0 \to \epsilon^{1/2}\,r_0 \quad , \quad k \to \epsilon^{1/2}k
  \quad , \quad c_2\to \frac{c_2}{\epsilon} \quad\quad
  \epsilon\to 0
\end{equation}
which can also be formulated in terms of $c_1$, by $c_1\to -\frac{8}{\sqrt{7}}
+ \epsilon\frac{k^2}{\sqrt{7}}$.

The previous scaling limit is certainly not the only possibility, and as it
will turn out important for us later on, we discuss a second possibility. 
Consider the following double scaling limit

\begin{equation}
  r_0\to \epsilon\,r_0 \quad , \quad c_2\to \frac{c_2}{\epsilon} \quad
  , \quad \epsilon\to 0 \quad [c_1\neq -\frac{8}{\sqrt{7}} \,\,\text{fixed}]
 \label{bpslimit}
\end{equation} 

It is clear that the charge \eqref{charge} remains finite in the limit 
\eqref{bpslimit} and that $\delta M$ vanishes, as required. As a further
check, it is straightforward to analyse \eqref{setfunctions} in the above 
limit to get back the BPS metric ~\cite{hs91} from \eqref{ansatzd6}.

\subsection{M-theory lift}

By rescaling the Einstein metric to the string frame and using the standard
Kaluza-Klein ansatz, one derives a family of purely geometrical configurations
in eleven dimensions described by the metric

\begin{multline}
  g =  \left(\frac{f_-(r)}{f_+(r)}\right)^{-c_1/6}ds^2(\bE^{1,6}) \\
  + \left(\frac{f_-(r)}{f_+(r)}\right)^{7c_1/12}\left(f_-(r)f_+(r)
  \right)^2\left[\cosh(k h(r)) - c_2 \sinh(k h(r))\right]
  \left(dr^2 + r^2d\Omega_2^2\right) \\
  + \left(\frac{f_-(r)}{f_+(r)}\right)^{7c_1/12}\left[
  \cosh(k h(r)) - c_2 \sinh(k h(r))\right]^{-1}\left(dz + C_1\right)^2 
 \label{msolution}
\end{multline}  

where $z$ stands for the spacelike coordinate along the M-theory circle with 
length at infinity $2 \pi g_s l_s$ and $C_{(1)}$ is the magnetic dual one 
form to the previous RR 7-form $[dC_{(1)}=\star_{10}dC_{(7)}]$.

Notice that whenever $c_1\neq 0$, the eleven dimensional geometry
is not that of seven dimensional Minkowski spacetime times some curved
manifold, but contains a warped factor. In the limit $r_0\to 0$ keeping
$c_1\,,c_2$ fixed, the geometry asymptotes to the maximally supersymmetric
Minkowski spacetime. 

One non-trivial check ~\cite{bmo00} for the above family of solutions
\eqref{msolution} concerns the zero charge sector $(Q=0)$. Indeed, it has
been known for a while the embedding in eleven dimensions ~\cite{s97} of
the Kaluza-Klein dipole solution ~\cite{gp83} describing a 
monopole-antimonopole pair separated by some distance. Studying such a 
solution in the limit of vanishing dipole size, one gets the
configuration

\begin{equation}
  g = ds^2(\bE^{1,6}) + r^2\left(\Delta^{-1}(r)dr^2 +d\Omega^2_2\right) +
  \Delta(r) r^{-2} dx^2\,,
 \label{sen1}
\end{equation} 
where the scalar function $\Delta (r)$ is defined by 
$\Delta(r)=r(r-2M)$, M being some constant parameter.

It is clear that the matching between \eqref{msolution} and \eqref{sen1}
requires setting
\[
  c_1=0 \quad , \quad c_2=1
\]
to ensure the vanishing of the warped factor and charge, respectively. 
The same reasoning applies for a system of more than two monopoles. If we 
want the solution to remain as a seven dimensional Minkowski spacetime 
times some four dimensional manifold where the monopoles are living, one 
needs $c_1 = 0$. In this subspace , \eqref{msolution} becomes

\begin{equation}
  g = ds^2(\bE^{1,6}) + \left(1 + \frac{r_0}{r}\right)^4 
  (dr^2 + r^2 d\Omega^2_2) + \left(\frac{r-r_0}{r+r_0} \right)^2 dx^2 ~.
 \label{nd6nad6}
\end{equation}

Notice that \eqref{sen1} and \eqref{nd6nad6} are equivalent, as expected, under
the coordinate transformation :
\[
  r = \hat{r}\left(f_+(\hat{r})\right)^2 \,,
\]
where $\hat{r}$ stands for the radial coordinate in \eqref{nd6nad6}, provided
the two constant parameters are identified as
\[
  M = 2\hat{r}_0\,.
\]

Notice that the right hand side of the above coordinate transformation
is invariant under the transformation $r_0/\hat{r}\to \hat{r}/r_0$.
We shall see later that this symmetry is not restricted to the vanishing
charge sector $(c_2=1)$, but generalizes to $Q\neq 0$.

\subsection{Two parameter solution in M-theory}

In the following, we shall concentrate on the $c_1=0\,[k=2]$ subspace
of solutions ~\cite{lt01}

\begin{eqnarray} 
  g &=& ds^2(\bE^{1,6}) + \left[\frac{1 - c}{2} f^4_- + \frac{1 + c}{2} f^4_+
  \right](dr^2 + r^2 d\Omega^2_2) 
  + \frac{(f_+ f_-)^2}{\left[\frac{1 - c}{2} f^4_- + 
  \frac{1 + c}{2} f^4_+\right]} (dz + C_{(1)})^2 \nonumber \\
 \label{msolution1}
\end{eqnarray}

which includes \eqref{nd6nad6} in the sector of zero charge $[c_2\equiv c=1]$.
We would like to emphasise that such a subspace of configurations includes
both the BPS ones, through the scaling limit \eqref{bpslimit}, and the zero
distance monopole-antimonopole pair solution \eqref{sen1}. Since it
contains a seven dimensional Minkowski spacetime, it allows us to concentrate
on the physics of the four dimensional curved manifold, which is rather
natural if one is interested in relating the physics of D6-$\overline{\text{D}6}$
at strong coupling with tachyon condensation in orbifold models in
$\bC^2$, whose local description close to the fixed point (singularity) 
consists of such a seven dimensional Minkowski spacetime times some four 
dimensional manifold.

The two parameters $\{r_0\,,c\}$ appearing in \eqref{msolution1} can be 
mapped to the charge \eqref{charge} and mass \eqref{mass} of the system, 
which satisfy the quadratic relation :

\begin{equation} 
  M^2 = Q^2 + (4\text{P}\cdot r_0)^2 ~,
\end{equation}

showing that the mass is bigger or equal to the charge. These parameters
can be expressed in a much more physical way in terms of the number of branes 
(N) and anti-branes $(\overline{\text{N}})$ as

\begin{equation}
  \begin{aligned}
    N-\bar{N} & = \frac{8}{g_s l_s} r_0\,\, \sqrt{c^2 -1}  \\
    N+\bar{N} & = \frac{8}{g_s l_s} r_0 c ~, 
  \end{aligned}
\end{equation}
or equivalently, by

\begin{equation}
  \begin{aligned}
    r_0^2 & = \frac{(g_s l_s)^2}{16} \,\, \ N \bar{N}  \\
    c & = \frac{N + \bar{N}}{2 \sqrt{N \bar{N}}} ~. 
  \end{aligned}
\end{equation}

As we can see from these formulae the radius of the bolt and the value of $c$ are discrete, as only an integer number of branes is allowed.

Notice that the measure for the non-BPS character of the configuration 
\eqref{nonbps} is proportional to the ratio 

\begin{equation}
  \delta M \propto V_6 \cdot \frac{R_s\cdot r_0}{l_p^9} ~,
\end{equation}

where $R_s$ is the radius of the M-theory circle and $l_p$ is the eleven
dimensional Planck length. A natural way of measuring the non-BPS character 
of the configuration in terms of D6-branes data is by the quotient

\begin{equation}
  \frac{N+\bar{N}}{N-\bar{N}} = \frac{c}{\sqrt{c^2 -1}}
\end{equation}
 
If there are only branes or antibranes, the quotient equals $\pm 1$, which
can only be achieved if $c \to \infty$. Notice that to keep the charge 
\eqref{charge} fixed in that limit, one must take at the same time 
$r_0 \to 0$, which matches our discussion on BPS limits, in particular
the scaling limit \eqref{bpslimit}.

As we shall discuss more extensively in the next section, there is a  
{\sl bolt} type singularity at $r_0$, both in the charged and non-charged
sectors, for non-zero values of $r_0$. When approaching the supersymmetric 
configuration, the fate of the {\sl bolt} singularity depends on the
sector in which we are :

\begin{itemize}
  \item[(i)] If $Q\neq 0$, it gives rise to the usual {\sl nut} singularity
  at $r=0$ where the monopoles (or antimonopoles) are sitting. This is the
  source for the naked singularity of the D6-branes (or 
  $\overline{\text{D}6}$-branes) at the origin ~\cite{p00}.
  \item[(ii)] If $Q=0$, it gives rise to flat space.
\end{itemize}

\section{Geometry of the solution}

Let us analyse the geometry of solution \eqref{msolution1}. First of all,
it is exactly the Taub-bolt singularity without imposing the absence
of conical singularities \cite{p78,lt01}. That can be seen explicitly
by the change of radial coordinate \cite{lt01} :

\begin{equation*}
  r^\prime = \frac{1}{2}\left(r-m + \sqrt{r^2 - 2m\,r + l^2}\right)~,
\end{equation*}

and identifying the parameters in both solutions as $c=m/\sqrt{m^2 - l^2}$
and $r_0=\sqrt{m^2-l^2}/2$. 

If we keep the charge fixed and take $r_0 \to 0$, or equivalently, we take 
the double scaling limit \eqref{bpslimit}, we end up with the Taub-NUT metric :

\begin{equation}
  g_4 = H(r) (dr^2 + r^2 d\Omega^2_2) + H(r)^{-1} (dz + C_{(1)})^2
 \label{taubnut}
\end{equation}

where $H(r)= 1 + 4 r_0 c/r$, as expected for the BPS configuration
(M=Q). In the limit close to the origin,  the metric \eqref{taubnut}
reproduces the singularity of a $\bZ_{\text{N}}$ orbifold 
($\text{A}_{\text{N}-1}$ singularity), where N is the number of branes 
defined previously, i.e. in D-brane units $H(r)= 1 + \frac{1}{2} g_s l_s 
\text{N}/r$. Indeed, close to the singularity located at $r=0$, one can make 
the coordinate transformation

\[
  \hat{r}= 2\left(\frac{1}{2} g_sl_s \text{N}\cdot r\right)^{1/2}~,
\]
which allows us to write the metric as

\begin{equation}
  g_4 = d\hat{r}^2 + \frac{\hat{r}^2}{4}\left[d\theta^2 + (\sin\theta)^2
  d\varphi^2 + \left(\frac{2 dz}{l_s g N} + 
  (1-\cos\theta)d\varphi\right)^2\right]~.
\end{equation}  

Taking into account that $z$ has a period of $2 \pi g_sl_s$ one gets that 
the circle parametrised by $z$ has a conical behaviour like a 
$\bZ_{\text{N}}$ orbifold.

The solution \eqref{msolution1} is defined for $r\geq r_0$, the interior
of the sphere $r=r_0$ not belonging to the solution. However, it is
interesting to point out the existence of an isometry, the in$\&$out
symmetry, that relates $r\ll r_0$ with $r\gg r_0$, 

\begin{equation*}
  \frac{r_0}{r} \to \frac{r}{r_0}~.
\end{equation*}

The geometry far away from $r\sim r_0$ has the same assymptotic behaviour as 
in the supersymmetric configuration. Thus, any source of instability 
reflected in the geometry has to be in the region $r\sim r_0$, at which we 
shall now look in detail.

Let us start our analysis in the charged sector $(Q\neq 0)$. Whenever
the configuration is non-BPS, the metric has a {\sl bolt} singularity
at $r=r_0$. The bolt is a sphere of radius proportional to $r_0$ with conical
singularities on it. To study these singularities, we can examine the metric
\eqref{msolution1} close to the bolt, by introducing the distance to the bolt
as a coordinate $(y=r-r_0)$ and concentrating on the region $y\ll r_0$.
After a trivial rescaling of the new radial coordinate, the four dimensional
metric looks like

\begin{equation} 
  g_4 = dy^2 + 8(1 + c) r_0^2 d\Omega^2_2 + \frac{y^2}{16(1 + c)^2 r_0^2} 
  (dz + C_{(1)})^2 ~.
 \label{mbolt}
\end{equation} 

Thus, close to the bolt, the periodicity of the compact coordinate $x=z/l_s$
is reduced by a factor

\begin{equation}
  \frac{1}{L} \equiv \frac{l_s g_s}{4 (1 + c) r_0} = 
  \frac{2}{N + \bar{N} + 2 \sqrt{N\bar{N}}} ~,
 \label{period}
\end{equation}

which indeed points out to the existence of conical singularities whose
angular deficit is $2 \pi \frac{L-1}{L}$. Notice that these singularities
are located on a sphere of radius $\text{R}_{\text{bolt}}=2\sqrt{2}\sqrt{1+c}\,
r_0$, whose area is

\begin{equation}
  A = 32 \pi (1+c)r_0^2 = \pi (l_s g_s)^2(N + \bar{N} + 2 \sqrt{N\bar{N}}) 
  \sqrt{N\bar{N}} ~.
\end{equation}

Notice that the area takes discrete values depending on the integer numbers 
representing the number of branes and antibranes.

Even though the scalar curvature vanishes on the bolt, due to the existence
of the conical singularities, one might wonder about higher order corrections 
to the eleven dimensional effective action close to the bolt. To clarify 
this issue, one can analyse the behaviour of the square of the Riemann tensor.
Such corrections would be suppressed whenever

\begin{equation*}
    l_p^4\,R_{MNPR}R^{MNPR} \ll 1 ~.
\end{equation*}

Working in the regime in which the number of branes is of the same order
as the number of antibranes $(N\sim\bar{N})$, the above constraint looks like

\begin{equation*}
  l_p^4\,R_{MNPR}R^{MNPR} \sim \left(\frac{l_p}{r_0}\right)^4\sim 
  \left(g_s^{2/3}\cdot N\right)^{-4} \ll 1 ~.
\end{equation*}

Therefore such corrections can be neglected when the size of the bolt is big
in eleven dimensional Planck units, or equivalently

\begin{equation}
  g_s^{2/3}\cdot N \gg 1 \quad , \quad N\sim \bar{N} ~.
 \label{aprox}
\end{equation}

Notice that in order to keep the string coupling constant small, the number 
of branes must be large. This is the approximation we would like to use.

When the size of the bolt is big $(r_0/l_p\gg 1)$, the metric \eqref{mbolt}
close to the bolt is a huge sphere times a cone. Furthermore, in the regime
\eqref{aprox}, the effect of $C_{(1)}$ is negligible \footnote{Globally the  structure of the space can be undertood as a $\bZ_L$ vector bundle over a trivial $\bZ_L$-space $S^2$. That means that the $\bZ_L$ is acting trivially on the sphere while rotating the fibre $\bC$. The charge $Q$ of the system specifies the first Chern number as in the supersymmetric case.}. Thus,
locally, the four dimensional manifold $\cM_4$ looks like

\[
  \cM_4 \sim \bC\times \bC/\bZ_{\text{L}} ~.
\]

That such a description allows an orbifold singularity $\bC/\bZ_{\text{L}}$
interpretation can be further checked by using \eqref{period} in the regime
\eqref{aprox}, which ensures that L is an integer number.

These orbifold singularities have always closed string tachyons in the
twisted sectors. In the next section, we shall compare the annihilation
of brane-antibrane pairs expected in the open string description, with the
sequences of transitions for $\bC/\bZ_{2\text{l}+1}$ orbifolds discussed
in ~\cite{aps01}, and we shall see that they are qualitatively the same.

\subsection{Same number of branes and antibranes}

We shall now move to the vanishing charge sector, that is, the one with the 
same number of branes and antibranes, i.e. $N=\bar{N}$. In this case, the 
metric reduces to \eqref{nd6nad6} and depends on a single parameter $r_0$,
which can be written in terms of the number $N$ of D6-$\overline{\text{D}6}$
pairs as

\[
  r_0 = \frac{1}{4}g_s l_s \cdot N ~.
\]

Since $C_{(1)}$ vanishes, the surfaces r=constant are trivial fibrations 
$S^1 \times S^2$. The assymptotic geometries are $\bR^{1,9}\times S^1$,
whereas close to $r\sim r_0$, one can check, proceeding in an analogous way
to the previous discussion, that the bolt structure remains. In this case,
the deficit in the periodicity is $1/2\text{N}$. That means that for an 
integer number of D6-branes the system has an orbifold 
interpretation as a $\bZ_{2N}$ orbifold. 

The scalar curvature vanishes everywhere, as it corresponds to a solution
of Einstein supergravity equations of motion with no matter, whereas the
squared of the Riemann tensor is given by

\begin{equation*}
  R_{MNPR}R^{MNPR} = 192 \frac{r^6 r_0^2}{(r + r_0)^{12}}~,
\end{equation*}

which has a maximum at $r=r_0$. Once more, the gravity approximation
is reliable in the large bolt limit.

\section{Evolution of the system}

When one trivially reduces the previous M-theory configurations 
\eqref{msolution1} by adding an extra transverse compact circle, one finds a 
generically non supersymmetric purely gravitational (geometrical) Type IIA 
configuration. Thus, the analysis of singularities discussed above still 
applies to this geometry. 

We are thus left with two different descriptions in type IIA of a single
M-theory configuration: first, the brane-antibrane system and on the other 
hand, geometrical configurations with conical singularities located on
a sphere. Furthermore, in the limit of big bolt \eqref{aprox}, the geometry 
of the conical singularities is locally given by that of an orbifold type,
$\bC\times\bC/\bZ_{\text{N}}$. Thus, it is clear that both systems contain 
tachyons; the brane-antibrane system in the open string sector from strings 
stretching between a brane and antibrane, whereas in the orbifold side,
there are closed string tachyons in the twisted sectors. These tachyons
can be understood as localised on the bolt. Some properties of this kind of 
closed string twisted sectors and their possible evolution have been analysed 
in ~\cite{aps01,v01,dv01,hkmm01}. In the following, we shall show that
the expected annihilation of brane-antibrane pairs in the open string
side matches the reduction in the order of the non-supersymmetric
orbifold observed in the previous cited references.

\begin{figure}
  \begin{center}
    \includegraphics{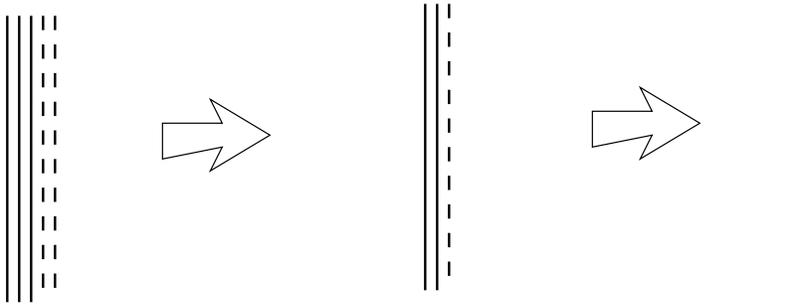}
    \caption{\small  Schematic representation of the D6 anti-D6-brane 
    annihilation to the supersymmetric configuration.}
   \label{brane}
  \end{center} 
\end{figure}

We can consider the evolution of the system in 
the $(M,Q)$ parameter space. In the D6-brane picture, we expect branes to 
annihilate the antibranes so that the total charge is preserved. The mass will 
decrease up to a supersymmetric system, $M=Q$, in which we are left
either with all branes or all antibranes. This process is expected to be a 
discontinuous process: branes and antibranes are annihilated in pairs as 
closed string fields will be emitted to the bulk. We expect a sequence 
\footnote{We call now M the number of branes plus antibranes, and Q its
difference.}
\[
  (M,Q) \rightarrow (M-2,Q) \rightarrow (M-4,Q) ... \rightarrow (Q,Q)
\] 
This process is represented schematically in figure ~\ref{brane}.

\begin{figure}
  \begin{center}
    \includegraphics{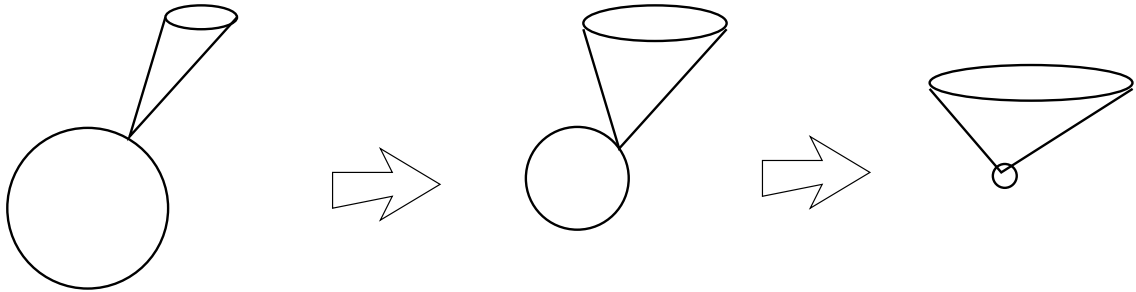}
    \caption{\small M-theory lift of the brane anti-brane annihilation 
    process. The two effects are the reduction of the bolt to a point and the 
    expanding of the cone to get a supersymmetric singularity 
    $\bC^2/\bZ_{\text{N}}$.}
   \label{bolt} 
  \end{center}
\end{figure}

\begin{figure}
  \begin{center}
    \includegraphics{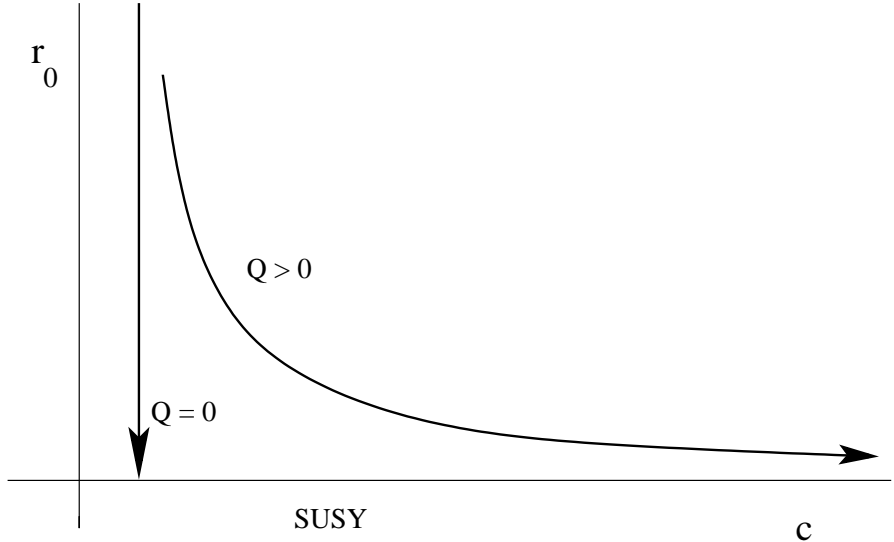}
    \caption{\small Flow in the $r_0$ and $c$ parameter space representing the annihilation of brane antibrane-pairs.}
   \label{parameter} 
  \end{center}
\end{figure}

When considered from the M-theory effective description in terms of a 
classical solution of the supergravity equations of motion, the latter
depends on two continuous parameters : $M$ and $Q$. Nevertheless, one can 
study the evolution in the geometry of the family of configurations by moving 
in such a two dimensional parameter space. Indeed, we are interested in 
studying the decrease in the mass $M$ while keeping the charge $Q$ fixed. It 
is clear that such a motion requires a decrease of $r_0$ while $c$ 
increases ``along the flow''. Heuristically, we can think of 
$M=N_1 + 2\bar{N}$ and $Q=N_1$ as the starting point of the flow. 
The value of $r_0$ is thus determined to be

\[
  r_0 = \frac{g_s l_s}{4} \sqrt{\bar{N}^2 + N_1\cdot\bar{N}}~.
\]
The motion along the flow we are interested in, is described by decreasing
the parameter $\bar{N}\to \bar{N}-2$, simulating the annihilation of a brane
and antibrane. One can formally take the limit $\bar{N}\to 0$ and get the
BPS configuration as expected. In the $r_0$, $c$ parameter space this flow 
can be seen as a curve going to $r_0 \to0$ and $c \to \infty$ 
(see \ref{parameter}).

This flow has two effects: the radius of the bolt goes to zero and the conical 
singularity gets 'less' conical with a factor $1/(M + \sqrt{M^2 -Q^2})$. When 
the system arrives at the supersymmetric configuration, the bolt disappears 
into a nut and a supersymmetric orbifold singularity remains at the origin 
$\bC^2/\bZ_{\text{Q}}$. See figure ~\ref{bolt}.

\begin{figure}
  \begin{center}
    \includegraphics{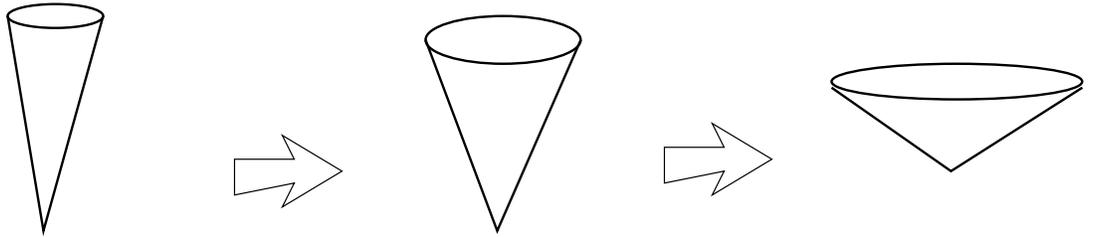}
    \caption{\small $\bC/\bZ_{\text{N}}$ orbifolds have always tachyons in 
    the closed string spectrum. By turning on some of this tachyons the cone 
    expands till reaching flat space.}
   \label{orbi}
  \end{center} 
\end{figure}

One very interesting case is when the number of branes $N$ is exactly the 
same as the number of antibranes $\bar{N}$. In this case, the flow corresponds
to a straight line at $c=1$, and the decrease in $r_0$ is directly related
to the decrease in $M=2N=r_0$. Then close to the bolt there is 
an orbifold description as $\bC/\bZ_{2N}$. The process of annihilating branes 
and antibranes takes $\text{N} \to \text{N}-2$. From the orbifold point of 
view that corresponds to a transition $\bC/\bZ_{2N} \to \bC/\bZ_{2(N -1)}$. 
Notice that this process is 
very similar to the one found by ~\cite{aps01} where the orbifold singularity 
is desingularising till reaches the flat space by $\bC/\bZ_{\text{2N+1}} 
\to \bC/\bZ_{\text{2N-1}}$ (see figure ~\ref{orbi}). Notice
that in the orbifold description in ~\cite{aps01}, the order of the
orbifold is odd while in our case is even. However as we have already
said, the correspondence between the two systems is expected to happen
only at large N.

Notice that in both sides, brane-antibrane annihilation and the vev of the 
twisted field are discontinuous, so our approximation of continuous mass 
variation has no meaning between these points. When interpreted in terms of 
branes and antibranes, we have seen that the radius of the bolt takes 
discrete values as well as the $c$ parameter. Notice that, as discussed in 
~\cite{aps01}, the process of desingularising the cone is expected to be 
discontinuous. So one expects sudden changes in the volume of the bolt from 
both sides. For example, one can consider the emission of dilaton fields by 
the brane-antibrane annihilation into the bulk. That will correspond to a 
sudden change in the M-theory coordinate that looks like the cone change 
in the twisted orbifold side as described in \cite{aps01}.
It will be very interesting to relate these two discontinuous processes in 
detail.  From the M-theory point of view, we can see the bolt as emitting 
waves that change suddenly the shape of the cone till the bolt disappear to a 
point.

It is important to notice that we are not mapping open string to closed 
string tachyons, we are just comparing the behaviour and evolution of two 
different systems related by an M-theory lift. If one naively tries to map 
one open to one closed string tachyon, one immediately realises that things 
are not working. For large number of pairs of branes and antibranes N the 
counting of open string tachyons goes like $\text{N}^2$ but the number of 
twisted closed string tachyons grows like N. Also the perturbative masses of 
these states do not match. However, the number of steps driving the system to 
the supersymmetric configuration is the same, of order N. This is because
when a pair brane-antibrane disappears there are also N open string tachyons 
that decouple from the spectrum.

\section{From orbifolds to branes}

The relation among $\bC/\bZ_{\text{N}}$ orbifolds and D6-$\overline{\text{D}6}$
systems in the large bolt limit leads us to consider a non-supersymmetric 
orbifold of Type IIA, perform its trivial lift to M-theory and reduce it
afterwards along a circle inside the orbifold. The configuration thus obtained
cannot be trusted far away from the origin, but it must correspond to the 
local description of some D6-brane system
\footnote{That is similar to what is happenning in flux-branes, 
see for instance, the discussion on  Ref. ~\cite{u01}.}. 
Notice that this is exactly what happens for the supersymmetric orbifold 
$\bC^2/\bZ_{\text{N}(\pm 1)}$ : this produces the familiar supersymmetric 
$\text{A}_{\text{N}-1}$ orbifolds (for review see ~\cite{abfgz93,j00}), as
reviewed at the beginning of section 3, which upon reduction along the Hopf
fibre, gives rise to the local description of a system of N coincident
D6-branes located at the fixed points of the $S^1$ along which we performed 
the reduction.

We shall next consider some particular non-supersymmetric orbifold 
singularities of the form 
$\bC^2/\left(\bZ_{\text{N}} \times \bZ_{\text{M}}\right)$, where the action
of each subgroup is defined in such a way that the complete orbifold
breaks supersymmetry completely. We will see that after reduction along
the Hopf fibre, the type IIA configuration has a line of conical 
singularities with some fractional D6-branes located at the origin
whenever $\text{M}\neq 0$. It is important to stress, once more, that the
forthcoming analysis is only reliable close to where the D-Branes are
located.

\subsection{$\bC^2/(\bZ_{\text{N}} \times \bZ_{\text{M}})$ orbifolds} 

Let us define polar coordinates in $\bC^2$ by
\begin{equation*}
  \begin{aligned}
    z_1 &= r\cos\frac{\theta}{2}\,e^{i(\psi + \varphi)/2} \\
    z_2 &= r\sin\frac{\theta}{2}\,e^{i(\psi - \varphi)/2} ~,
  \end{aligned}
\end{equation*}
where the range of the different angular variables is
$0\leq \theta < \pi$, $0\leq \varphi < 2\pi$ and $0\leq \psi < 4\pi$. 
The action of the $\bZ_{\text{N}} \times \bZ_{\text{M}}$ group on $\bC^2$ 
is of the form :

\begin{equation}
  g_1(z) = \left(
  \begin{array}{cc}
    e^{\frac{2 \pi i}{N}} & 0 \\ 
    0 &  e^{\frac{2 \pi i}{N}}
  \end{array}
  \right) 
  \left( 
  \begin{array}{c}
    z_1 \\ z_2
  \end{array} 
  \right)
\end{equation}

and

\begin{equation}
  g_2(z) = \left(
  \begin{array}{cc}
    e^{\frac{2 \pi i}{M}} & 0 \\ 
    0 &  e^{\frac{- 2 \pi i}{M}}
  \end{array}
  \right) 
  \left( 
  \begin{array}{c}
    z_1 \\ z_2
  \end{array} 
  \right)
\end{equation}

where $g_i$ are the generators of the group. The orbifold does not preserve
supersymmetry because each subgroup preserves supersymmetries of different
chirality. Thus, whenever the order of both subgroups (N,M) is different 
from 1, the total orbifold breaks supersymmetry completely. Due to the
identifications associated with the orbifold construction, there are
now two cones associated with each of the subgroups 
\[
  0\leq \varphi < \frac{2\pi}{M} \quad \text{and} \quad 
  0\leq \psi < \frac{4\pi}{N}~.
\]

One can work with angular variables satisfying the standard periodicity
conditions by rescaling $\{\varphi\,,\psi\}$. In this way, the periods
are manifest in the metric

\begin{equation}
  ds^2 = ds^2(\bE^{1,6}) + dr^2 + \frac{r^2}{4} \left[
  d\theta^2 + sin^2\theta \frac{d\varphi^2}{M^2} + 
  \left( \frac{d\psi}{\text{N}} + cos\theta \frac{d\varphi}{\text{M}}
  \right)^2\right]
\end{equation}

There are many $S^1$'s along which one could reduce, but we shall take 
the usual Hopf fibering, i.e. reducing on $\psi$. Using the Kaluza-Klein
ansatz, the ten dimensional metric in the string frame looks like

\begin{equation}
  ds^2 = \frac{r}{2\text{N}}\left\{ds^2(\bE^{1,6}) + dr^2 + \frac{r^2}{4}
  \left(d\theta^2 + (\sin\theta)^2 \frac{d\varphi^2}{\text{M}^2}
  \right)\right\}~, 
\end{equation}

whereas the dilaton and RR one form are given by:

\begin{equation}
  \begin{aligned}
    e^{\Phi} & = \left(\frac{r}{2\text{N}}\right)^{\frac{3}{2}} \\
    C_{(1)} & = \frac{\text{N}}{\text{M}} \cos\theta d\varphi
  \end{aligned}
\end{equation}

\begin{figure}
  \begin{center}
    \includegraphics{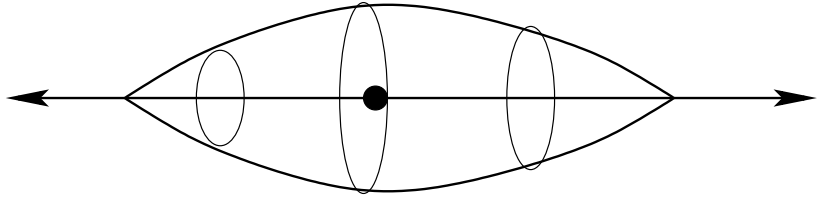}
    \caption{\small Reduction of a non-supersymmetric 
    $\bC^2/(\bZ_{\text{N}} \times \bZ_{\text{M}})$ orbifold. At a fix 
    distance from the origin where the fractional D6-brane is located the $S^2$ presents two conical singularities that represents the intersection of the two dimensional sphere with a line of $\bC/\bZ_{\text{M}}$ singularities.}
   \label{NM}
  \end{center} 
\end{figure}

Notice that if $M=1$, the above configuration matches the local description
of N coincident D6-branes close to the naked singularity (r=0), and half
of the supersymmetry is preserved.

Whenever $\text{M}\neq 1$, the naked singularity remains but there is
an additional line of conical singularities coming from a $\bC/\bZ_{\text{M}}$
orbifold. Indeed, after reducing along the Hopf fibering, we are left with
$\bR^3$ in the subspace transverse to the D6-branes, but with one angular
coordinate of reduced period \footnote{Using $\varphi^\prime=\varphi/M$,
the metric is flat but $\varphi^\prime$ is defined over $0\leq \varphi^\prime
< 2\pi/M$.}. The set of fixed points of the orbifold which reduced the period
of the angular variable is given by the line $\theta=0\,,\pi \,\forall$ r.
We can thus interpret the ten dimensional configuration as the local 
description close to r=0 of a set of D-branes on a $\bC/\bZ_{\text{M}}$
orbifold carrying fractional charge N/M.

Notice that for the systems just discussed there is no open-closed string 
instability correspondence like in previous sections, since the analysis in 
both sides implies the existence of closed string tachyons in the twisted 
sectors. 

\section{Conclusions and perspectives}

In this paper, we have analysed the geometry of the lift to M-theory of
certain D6-$\overline{\text{D}6}$ systems. For any non-BPS configuration,
we find a bolt type singularity. The annihilation of D6-$\overline{\text{D}6}$
pairs in the open string description is realised, on the gravity side, by a
reduction on the size of the bolt and a desingularization of the conical
singularities on it. In the large bolt limit, the M-theory geometry is locally
described by $\bC\times \bC/\bZ_{\text{N}}$. This allowed us to qualitatively
match the annihilation of D6-$\overline{\text{D}6}$ pairs with the sequences
of transitions described in $\bC/\bZ_{\text{N}}$
non-supersymmetric orbifolds.  As we have already said, the process is 
discontinuous in both sides. It would be very interesting to analyse how the 
discrete evolution is produced. Having realised this connection, we considered
the non-supersymmetric orbifold $\bC^2/\bZ_{\text{N}}\times\bZ_{\text{M}}$
and its relation with a local description of unstable branes, which turned
out to be fractional D6-branes on a $\bC/\bZ_{\text{M}}$ singularity.

There are several natural questions related with the results reported here.
Due to the relation among D6-brane systems and $\bC/\bZ_{\text{N}}$ orbifolds,
it would be very interesting to investigate if there is any brane realisation
for the sequences of transitions found in ~\cite{aps01} regarding 
non-supersymmetric $\bC^2/\bZ_{\text{N(k)}}$ orbifolds.

We would also like to point out that the brane-antibrane system discussed
in this paper can be interpreted as a particular case of a pair of
D6-branes at generic angles, the one in which they have opposite
orientations. These more general systems do generically break supersymmetry
\footnote{See, among others, \cite{angle}.} and in some regions of their 
moduli space, they are empty of tachyons. It would be interesting to 
understand the M-theory dynamics in these cases ~\cite{angel}.

Other physical systems which have recently been given a lot of attention
and do also have localised closed string tachyons are fluxbranes
~\cite{flux}. It would be interesting to understand the stability and 
supersymmetry of some of them using similar local 
descriptions to the ones appearing in this work.

On the other hand, the analysis in section 5 is just a local one, as can be
seen from the fact that the dilaton (string coupling) increases as we move away
from the origin. It would be nice to look for non-BPS configurations whose
validity of description goes beyond the region where the brane sits.

\medskip
\section*{Acknowledgments}
\noindent
We would like to thank P. Barb\'on, R. Emparan, Y. Oz and A. Uranga for 
discussions, and especially R. Emparan for pointing out the existence of
reference \cite{lt01} and for his comments on the first version of this
work. R.R. would like to thank the group in Humboldt University in 
Berlin for hospitality during the progress of this work. J.S. would like to 
thank the theory division at CERN for hospitality during the initial stages
of the present work. The research of J. S. has been supported by a Marie Curie 
Fellowship of the European Community programme ``Improving the Human 
Research Potential and the Socio-Economic knowledge Base'' under the 
contract number HPMF-CT-2000-00480.


\end{document}